\documentclass[usegraphicx,usenatbib]{mn2e}
\renewcommand\[{\begin{equation}}
\renewcommand\]{\end{equation}}

\def\xv{{\bf x}}
\def\yv{{\bf y}}
\def\av{{\bf a}}
\def\xci{x_i}
\def\yci{y_i}
\def\Phic{\Phi_c}
\def\Psic{\Psi_c}
\def\phieff{\Phi_{\rm eff}}
\def\rhoc{\rho_c}
\def\zetac{\zeta_c}
\def\rc{r_c}
\def\vel{v_{\varphi}}
\def\velq{\vel^2}

\catcode`\@=11
\def\gsim{\ifmmode{\mathrel{\mathpalette\@versim>}}
    \else{$\mathrel{\mathpalette\@versim>}$}\fi}
\def\lsim{\ifmmode{\mathrel{\mathpalette\@versim<}}
    \else{$\mathrel{\mathpalette\@versim<}$}\fi}
\def\@versim#1#2{\lower 2.9truept \vbox{\baselineskip 0pt \lineskip
    0.5truept \ialign{$\m@th#1\hfil##\hfil$\crcr#2\crcr\sim\crcr}}}
\catcode`\@=12

   \title[Exact density-potential pairs]{Exact density-potential pairs 
          from the holomorphic Coulomb field}

   \author[Ciotti \& Giampieri]
          {Luca Ciotti$^1$ and Giacomo Giampieri$^2$
           \\ $^1$Astronomy Department, University of Bologna, 
                       via Ranzani 1, 40127 Bologna, Italy
           \\ $^2$Jet Propulsion Laboratory, California Institute of 
                  Technology, Pasadena CA 91109, USA
          }

\date{Accepted version, January 12, 2007}
\pubyear{2007}

\begin{document} 
\maketitle

\begin{abstract} 

We show how the complex-shift method introduced by Appell in gravity
to the case of a point mass (and applied among others in
electrodynamics by Newman, Carter, Lynden-Bell, and Kaiser to
determine remarkable properties of the electromagnetic field of
rotating charged configurations), can be extended to obtain new and
explicit density-potential pairs for self-gravitating systems
departing significantly from spherical symmetry.  The rotational
properties of two axysimmetric baroclinic gaseous configurations
derived with the proposed method are illustrated.

\end{abstract}

\begin{keywords}
celestial mechanics -- stellar dynamics -- galaxies: kinematics and dynamics
\end{keywords}

\section{Introduction}

For the discussion of many astrophysical problems where gravity is
important, a major difficulty is set by the potential theory. In
general, to calculate the gravitational potential associated with a
given density distribution one has to evaluate a three-dimensional
integral. Except for special circumstances, where a solution can be
found in terms of elementary functions, one has to resort to numerical
techniques and sophisticated tools, such as expansions in orthogonal
functions or integral transforms.

Under spherical symmetry, the density-potential relation can be
reduced to a one-dimensional integral, while for axisymmetric systems
one in general is left with a (usually non-trivial) two-dimensional
integral. As a result, the majority of the available explicit
density-potential pairs refers to spherical symmetry and only few
axially symmetric pairs are known (e.g., see Binney \& Tremaine 1987,
hereafter BT).  In special cases (in particular, when a
density-potential pair can be expressed in a suitable parametric form)
there exist systematic procedures to generate new non-trivial
density-potential pairs (e.g., see the case of Miyamoto \& Nagai 1975
and the related Satoh 1980 disks; see also Evans \& de Zeeuw 1992; de
Zeeuw \& Carollo 1996).  For non-axisymmetric systems the situation is
worse. One class of triaxial density distributions for which the
potential can be expressed in a tractable integral form is that of the
stratified homeoids, such as the Ferrers (1887) distributions (e.g.,
see Pfenniger 1984, Lanzoni \& Ciotti 2003, and references therein)
and special cases of the family considered by de Zeeuw \& Pfenniger
(1988). Additional explicit density-potential pairs are given by the
Evans (1994) models and by those constructed with the Kutuzov-Osipkov
method (1980, see also Kutuzov 1998).  Another way to construct in a
systematic way explicit density-potential pairs with finite deviations
from spherical symmetry has been presented in Ciotti \& Bertin
(2005). This technique is based on an elementary property of the
asymptotic expansion of the homeoidal potential quadrature formula for
small flattenings\footnote{Such expansion can be traced back to the treatise
on geodesy by Sir H. Jeffreys (1970, and references therein; see also
Hunter 1977).}, and recently, it has been applied to the modeling of gaseous
halos in clusters (Lee \& Suto 2003, 2004), to the study of the
dynamics of elliptical galaxies (Muccione \& Ciotti 2003, 2004), and
finally to a possible interpretation of the rotational field of the
extraplanar gas in disk galaxies (Barnab\`e et al. 2005, 2006).

In the context of classical electrodynamics a truly remarkable and
seamingly unrelated result has been obtained by Newman (1973, see also
Newman \& Janis 1965, Newman et al. 1965), who considered the case of
the electromagnetic field of a point charge displaced on the imaginary
axis. Successive analysis of Carter (1968), Lynden-Bell (2000, 2002,
2004ab, and references therein; see also Teukolsky, 1973;
Chandrasekhar, 1976; Page, 1976), and Kaiser (2004, and references
therein) revealed the astonishing properties of the resulting
holomorphic "magic" field.  As pointed out to us by the Referee, the
complex-shift method was first introduced by Appell (1887, see also
Whittaker \& Watson 1950) to the case of the gravitational field of a
point mass, and successively used in General Relativity (e.g., see
Gleiser \& Pullin 1989; Letelier \& Oliveira 1987, 1998; D'Afonseca,
Letelier \& Oliveira, 2005 and references therein).

In this paper we show how the complex-shift method can be also applied
in classical gravitation to obtain exact and explicit
density-potential pairs deviating significantly from spherical
simmetry; we are not aware that this possibility has been noticed in
the literature. The paper is organized as follows.  In Sect. 2 we
present the idea behind the method, and we discuss its application to
the case of spherical systems, describing two new self-gravitating and
axisymmetric systems obtained from the Plummer and Isochrone spheres.
In Sect. 3 we focus on the rotational fields of the new pairs when
interpreted as gaseous systems. Section 4 summarizes the main results
obtained, while in the Appendix we present the peculiar behavior of
the singular isothermal sphere under the action of the complex shift.

\section{General considerations}

We start by extending the complexification of a point charge Coulomb
field discussed by Lynden-Bell (2004b), to the gravitational potential
$\Phi(\xv)$ generated by a density distribution $\rho(\xv)$. From now
on $\xv =(x,y,z)$ will indicate the position vector, while
$<\xv,\yv>\equiv\xci\yci$ is the standard inner product over the reals
(repeated index summation convention implied).

Let assume that the (nowhere negative) density distribution $\rho
(\xv)$ satisfies the Poisson equation
\[
\nabla^2\Phi=4\pi G\rho,
\label{poisson}
\]
and that the associated complexified potential $\Phic$ with shift $i
\av$ is defined as
\[
\Phic(\xv)\equiv \Phi(\xv -i\av),
\]
where $i^2=-1$ is the imaginary unit and $\av$ is a real vector.  The
idea behind the proposed method is based on the recognition that
1) the Poisson equation is a linear PDE, and that 2) the complex shift
is a linear coordinate transformation. From these two properties, 
and from equations (1) and (2) it follows that
\[
\nabla^2\Phic=4\pi G\rhoc,
\]
where 
\[
\rhoc(\xv)\equiv \rho(\xv -i\av).
\]
Thus, by separating the real and imaginary parts of $\Phic$ and
$\rhoc$ obtained from the shift of a known real density-potential pair
one obtains {\it two} real density-potential pairs.

A distinction is in order here between electrostatic and gravitational
problems: in fact, while in the former case a density (charge)
distribution with negative and positive regions can be (at least
formally) accepted, in the gravitational case the obtained density
components have physical meaning only if they do not change sign (see
however Sect. 4). Quite interestingly, some general result about the
sign of the real and imaginary parts of the shifted density can be
obtained by considering the behavior of the complexified
self-gravitational energy and total mass. In fact, from the linearity
of the shift, it follows that the volume integral over the entire
space
\begin{eqnarray}
W_c&\equiv&{1\over 2}\int\rhoc\Phic d^3\xv\nonumber\\ 
&=&{1\over 2}\int [\Re(\rhoc)\Re(\Phic) - \Im(\rhoc)\Im(\Phic)] d^3\xv
\end{eqnarray}
coincides with the self-gravitational energy $W=0.5\int\rho\Phi
d^3\xv$ of the real unshifted seed density.  Thus the imaginary part
of $W_c$ is zero, i.e.
\begin{eqnarray}
{1\over 2}\int [\Re(\rhoc)\Im(\Phic)+ \Im(\rhoc)\Re(\Phic)] 
                 d^3\xv&=&\nonumber\\
          -G\int\int{\Re[\rhoc (\xv)]\Im[\rhoc(\xv ')]\over ||\xv-\xv'||} 
                 d^3\xv d^3\xv '&=&0,
\label{wcim}
\end{eqnarray}
and $W_c$ is the difference of the gravitational energies of the real
and the imaginary parts of the shifted density.  The vanishing of the
double integral (\ref{wcim}) shows that the integrand necessarily
changes sign, i.e., the complex shift cannot generate {\it two}
physically acceptable densities.  Additional informations are provided
by considering that, for the same reasons just illustrated, also the
total mass of the complexified distribution $M_c=\int\rhoc d^3\xv$
coincides with the total (real) mass of the seed density distribution
$M=\int\rho d^3\xv$, so that
\[
\int\Im(\rhoc) d^3\xv =0.
\label{imM}
\]
Identities~(\ref{wcim})-(\ref{imM}) then leave open the question
whether at least $\Re(\rhoc)$ can be characterized by a single sign
over the whole space.  It results that either cases can happen,
depending on the specific seed density and shift vector adopted. In
fact, in the following Section we show that simple seed densities
exist so that $\Re(\rhoc)$ is positive everywhere, while in the
Appendix we present a simple case in which also $\Re(\rhoc)$ changes
sign.

\subsection{Spherically symmetric ``seed'' potential}

We now restrict the previous general considerations to a spherically
symmetric real potential $\Phi(r)$, where $r\equiv
||\xv||=\sqrt{<\xv,\xv>}$ is the spherical radius, and $|| ... ||$ is
the standard Euclidean norm.  After the complex shift the norm must be
still interpreted over the reals\footnote{If one adopt the standard
inner product over the complex field, one would obtain $||\xv
-i\av||^2 = r^2 +||\av||^2$.} (e.g., see Lynden-Bell 2004b), so that
\[
||\xv -i\av||^2=r^2 -2i<\xv,\av> -a^2,
\]
where $a^2\equiv ||\av ||^2$.  Without loss of generality we assume
$\av=(0,0,a)$, and the shifted radial coordinate becomes
\[
\rc^2 = r^2-2iar\mu -a^2;\quad\quad \mu\equiv\cos\theta,
\]
where $\theta$ is the colatitude of the considered point, $\mu r=z$,
and
\[
\Phic=\Phi\left(\sqrt{r^2-2iaz -a^2}\right),
\label{eqpsic}
\]
\[
\rhoc=\rho\left(\sqrt{r^2-2iaz -a^2}\right).
\label{eqrhoc}
\]
Note that, when starting from a spherically symmetric seed system, the
real and immaginary parts of the shifted density $\rhoc$ can be
obtained 1) from evaluation of the Laplace operator applied to the
real and imaginary parts of the potential $\Phic$, 2) by expansion of
the complexified density in equation~(\ref{eqrhoc}), or finally 3) by
considering that for spherically symmetric systems $\rho=\rho(\Phi)$,
and so the real and imaginary parts of $\rhoc(\Phic)$ can be expressed
(at least in principle) as functions of the real and imaginary part of
the shifted potential.

\subsection{The shifted Plummer sphere} 

As a first example, in this Section we apply the complex shift to the
Plummer (1911) sphere.  We start from the relative potential
$\Psi=-\Phi$, where
\[
\Psi={GM\over b}{1\over\zeta},\quad\zeta\equiv\sqrt{1+r^2},
\label{plumpot}
\]
and $r$ is normalized to the model scale-lenght $b$, so that the
associated density distribution is
\[
\rho={3M\over 4\pi b^3}{1\over\zeta^5}
\]
(e.g., see BT).  For ease of notation, from now
on we will use normalized density (to $M/b^3$) and potential (to
$GM/b$), and so
\[
\rhoc={3\over 4\pi}[\Re(\Psic)+i\Im(\Psic)]^5.
\label{rhopcplu}
\]
From substitution (8) (where also the shift $a$ is expressed in $b$
units), the shifted potential $\Psic=1/\zetac$ depends on the square
root of $\zetac^2=1-a^2+r^2 -2iaz\equiv d{\rm e}^{i\varphi}$, with
\[
d\equiv |\zetac|^2=\sqrt{(1-a^2+r^2)^2 +4a^2z^2},
\]
and
\[
\cos\varphi={1-a^2+r^2\over d},\quad\sin\varphi=-{2az\over d}.
\]
Note that $\cos\varphi >0$ everywhere for $a<1$, and in
the following discussion we restrict to this case. The square
root 
\[
\zetac=\sqrt{d} {\rm e}^{\pi k i+\varphi i/2}, \quad\quad(k=0,1),
\label{zeta}
\] 
is made a single-valued function of $(r,z)$ by cutting the complex
plane along the negative real axis (which is never touched by
$\zetac^2$) and assuming $k=0$, so that the model equatorial plane is
mapped into the line $\varphi=0$. With this choice the principal
determination of $\Psic$ reduces to $\Psi$ when $a=0$, and simple
algebra shows that
\[
\cos{\varphi\over 2}={\sqrt{1+\cos\varphi}\over\sqrt{2}},\quad 
\sin{\varphi\over 2}=-{\sqrt{2}az\over d\sqrt{1+\cos\varphi}}.
\]
The real and imaginary parts of $\Psic$ are then given by 
\[
\Re (\Psic)=\Re\left({\bar\zetac\over |\zetac|^2}\right)
           ={\sqrt{d+1-a^2+r^2}\over \sqrt{2}\,d},
\label{plumre}
\]
\[ 
\Im (\Psic)=\Im\left({\bar\zetac\over |\zetac|^2}\right)
           ={az\over d^2\Re(\Psic)},
\label{plumim}
\] 
respectively, and from equation~(\ref{rhopcplu}) we obtain the
expressions of the (normalized) axysimmetric densities:
\[
\Re(\rhoc)={3\Re(\Psic)\over 4\pi}\left[
           \Re(\Psic)^4-{10 a^2z^2\over d^4}+
           {5a^4z^4\over d^8\Re(\Psic)^4}
           \right],
\label{plumrerho}
\]
\[
\Im(\rhoc)={3\Im(\Psic)\over 4\pi}\left[
           5\Re(\Psic)^4-{10a^2z^2\over d^4}+
           {a^4z^4\over d^8\Re(\Psic)^4}
           \right].
\label{plumimrho}
\]
We verified that the two new density-potential pairs
(\ref{plumre})-(\ref{plumrerho}) and (\ref{plumim})-(\ref{plumimrho})
satisfy the Poisson equation~(\ref{poisson}).

Note that, at variance with $\Re(\Psic)$, the potential $\Im(\Psic)$
change sign crossing the equatorial plane of the system. In accordance
with this change of sign, also $\Im(\rhoc)$ is negative for $z<0$, as
can be seen from equation~(\ref{plumimrho}), and thus cannot be used
to describe a gravitational system; we do not discuss this pair any
further, while we focus on the real components of the shifted
density-potential pair.  Near the center $d\sim 1-a^2
+r^2[1+2a^2\mu^2/(1-a^2)]+O(\mu^2 r^4)$, and the leading terms of the
asymptotic expansion of equations~(\ref{plumre}) and (\ref{plumrerho})
are
\[
\Re(\Psic)\sim {1\over\sqrt{1-a^2}}
               -{r^2(1-a^2+3a^2\mu^2)\over 2 (1-a^2)^{5/2}}
               +O(r^4),
\]
\[
\Re(\rhoc)\sim {3\over 4\pi (1-a^2)^{5/2}}
               -{15r^2 (1-a^2 +7a^2\mu^2)\over 8\pi (1-a^2)^{9/2}}
               +O(r^4);
\]
in particular, the isodenses are oblate ellipsoids with minor-to-major
squared axis ratio $(1-a^2)/(1+6a^2)$.  For $r\to\infty$, $d\sim r^2
+ 1-a^2+2a^2\mu^2 + O(\mu^2 r^{-2})$, and
\[
\Re(\Psic)\sim {1\over r}-{1-a^2+3a^2\mu^2\over 2 r^3}+O(r^{-5}),
\]
\[
\Re(\rhoc)\sim {15\over 4\pi r^5}
\left({1\over 5} - {1-a^2+7a^2\mu^2\over 2 r^2}\right)+O(r^{-9}),
\]
so that $\Re(\rhoc)$ coincides with the unshifted seed density (13)
and it is spherically symmetric and positive\footnote{This is a
general property of shifted spherical systems, as can be seen just by
expanding equation~(\ref{eqrhoc}) for $r\to\infty$.}.  Thus, near the
center and in the far field $\Re(\rhoc)>0$ for $0\leq a<1$.  In
addition, on the model equatorial plane $z=0$ (where $d=1-a^2+R^2$,
and $R$ is the cylindrical radius), $\Re(\Psic)$ coincides with the
potential of a Plummer sphere of scale-lenght $\sqrt{1-a^2}$, and from
equation~(\ref{plumrerho}) it follows that
$\Re(\rhoc)=3\Re(\Psic)^5/(4\pi)>0$ for $0\leq a<1$.

However, for $z\neq 0$ a negative term is present in
equation~(\ref{plumrerho}), and the positivity of $\Re(\rhoc)$ is not
guaranteed for a generic value of the shift parameter in the range
$0\leq a<1$. In fact, a numerical exploration reveals that
$\Re(\rhoc)$ becomes negative on the symmetry axis $R=0$ at $z\simeq
0.81$ for $a=a_m\simeq 0.588$; the negative density region then
expands around this critical point for increasing $a>a_m$. The
isodensity contours of $\Re(\rhoc)$ in a meridional plane are shown in
the top panels of Fig. 1, for shift parameter values $a=1/2$ (left)
and $a =23/40=0.575$ (right). The most salient property is the
resulting toroidal shape of the model with the large shift, which
reminds similar structures known in the literature, as the Lynden-Bell
(1962) flattened Plummer sphere, the densities associated with the
Binney (1981) logarithmic potential and the Evans (1994) scale-free
potentials, the Toomre (1982) and Ciotti \& Bertin (2005) tori (see
also Ciotti, Bertin, \& Londrillo 2004), and the exact MOND
density-potential pairs discussed in Ciotti, Nipoti \& Londrillo
(2006). Unfortunately, we were not able to find the explanation (if
any) behind this similarity (see however Sect. 4).

%%%%%%%%%%%%%%%%%%%%%%%%%%%%%%%%%%%%%%%%%%%
\begin{figure*}
\vskip -0.7truecm
\hskip -1.7truecm
\includegraphics[height=0.7\textheight,width=0.9\textwidth]{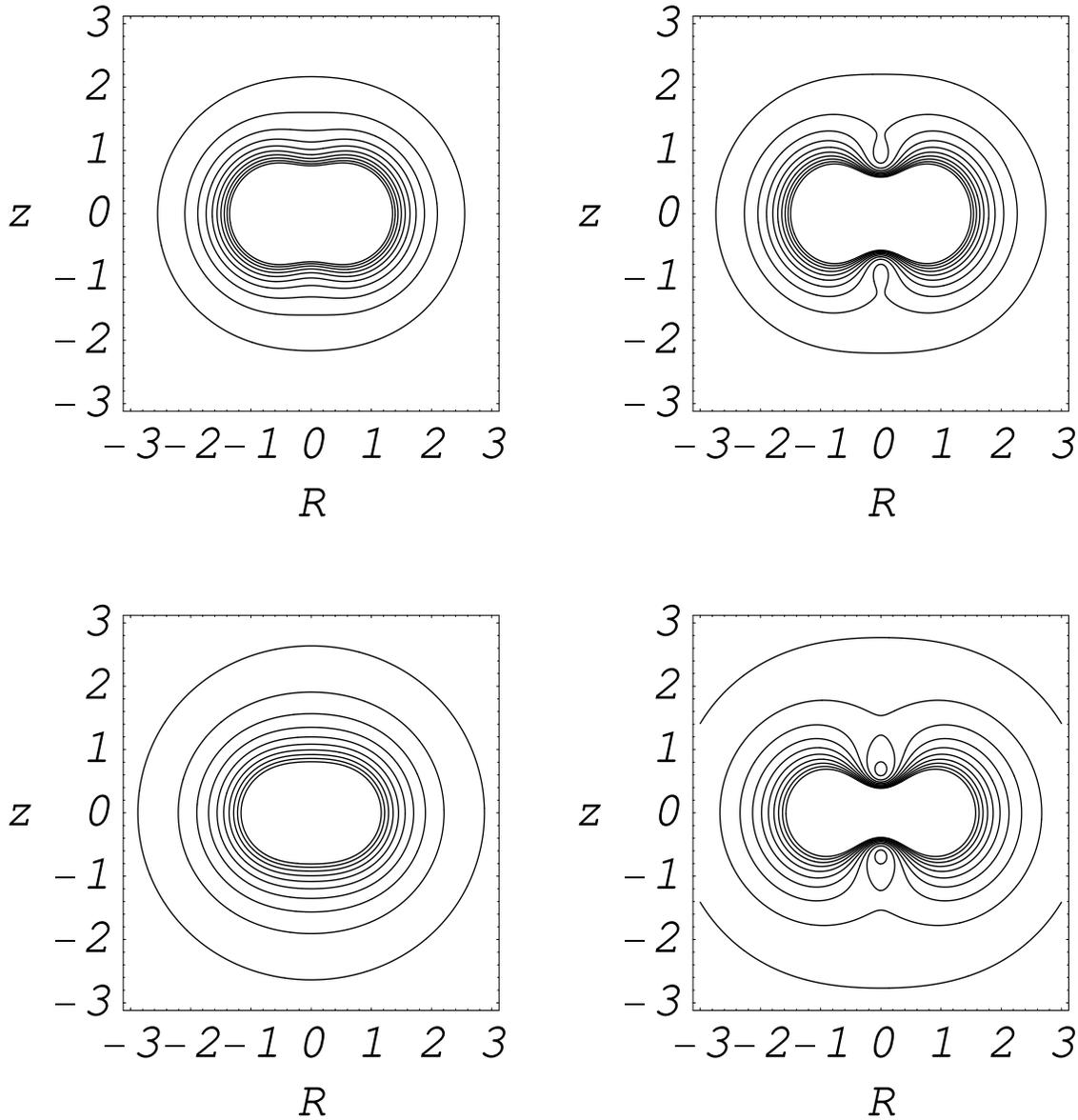}
\caption{Isodensity contours in the $(R,z)$ plane of $\Re(\rho_c)$ of
the shifted Plummer sphere for $a=1/2$ (top left) and $a=23/40$ (top
right), and of the shifted Isochrone sphere for $a=1/2$ (bottom left)
and $a=4/5$ (bottom right).  The coordinates are normalized to the
scale-lenght $b$ of the corresponding seed spherical model.}
\label{isodensity}
\end{figure*}
%%%%%%%%%%%%%%%%%%%%%%%%%%%%%%%%%%%%%%%%%%%

\subsection{The shifted Isochrone sphere} 

Following the treatment of the Plummer sphere, we now consider the
slightly more complicate case of the shifted Isochrone sphere. Its
relative potential and density are given by
\[
\Psi={GM\over b}{1\over 1+\zeta},
\]
\[
\rho={M\over 4\pi b^3}{3(1+\zeta)+2r^2\over
                   (1+\zeta)^3\zeta^3},
\]
where $\zeta$ is defined in equation (12), and again all lenghts are
normalized to the scale $b$ (H\'enon 1959, see also BT). In
analogy with equation (14), the normalized shifted density can be
written as a function of the normalized shifted potential
\[
\rhoc={\Psic^2\over 4\pi\zetac^3}\left(3+2\rc^2\Psic\right).
\]
The real and imaginary parts of 
$\Psic=(1+\bar\zetac)/[1+|\zetac|^2+2\Re(\zetac)]$ are easily obtained
from equations (15)-(18) as
%\[
%\Re(\Psic)={1+\sqrt{d}\cos(\varphi/2)\over 
%           1 +2\sqrt{d}\cos (\varphi/2)+d},
%\]
%\[
%\Im(\Psic)={\sqrt{2}az\over d\sqrt{1+\cos\varphi} 
%           [1+2\sqrt{d}\cos (\varphi/2) +d]}.
%\]
\[
\Re(\Psic)={1+\sqrt{d+1-a^2+r^2}/\sqrt{2}\over 
            1 +d+\sqrt{2(d+1-a^2+r^2)}},
\]
\[
\Im(\Psic)={\sqrt{2}az\over 
            \sqrt{d(d+1-a^2+r^2)} 
            [1+d+\sqrt{2(d+1-a^2+r^2)}]},
\]
while $\Re(\rhoc)$ and $\Im(\rhoc)$ can be obtained by expansion of
equation (29); however their expression is quite cumbersome, and so
not reported here.  Again, $\Im(\Psic)$ changes sign when crossing the
model equatorial plane, revealing that also $\Im(\rhoc)$ changes sign
in order to produce the resulting vertical force field near $z=0$, and
the vanishing of the integral in equation~(\ref{imM}).  In accordance
with equation~(\ref{eqrhoc}) on the model equatorial plane
$\Re(\rhoc)$ is functionally identical to an Isochrone sphere of
scale-lenght $\sqrt{1-a^2}$ (and so positive for $0\leq a<1$), while
for $r\to\infty$
\[
\Re(\Psic)\sim {r-1\over r^2}+{1+a^2-3a^2\mu^2\over 2 r^3}+O(r^{-4}),
\]
\[
\Re(\rhoc)\sim {1\over 4\pi r^5}
\left(2r-3 +4a^2{1-6\mu^2\over r}\right)+O(r^{-7}).
\]
As for the shifted Plummer sphere, also in this case $\Re(\rhoc)$
coincides in the far field with the seed density, and so it is
positive for $0\leq a<1$, while it becomes negative on the $z$ axis
at $z\simeq 0.648$ for $a>a_m\simeq 0.804$.  In the bottom panels of
Fig. 1 we show the isodensity contours in the meridional plane of
$\Re(\rhoc)$ for the representative values of the shift parameter
$a=1/2$ (left) and $a=4/5$ (right): again, the toroidal shape and the
critical regions on the symmetry axis are apparent in the case of the
larger shift.

\section{Rotational fields} 
 
A natural question to ask is about the nature of the kinematical
fields that can be supported by the found density-potential pairs
$\Re(\Psic)-\Re(\rhoc)$ when considered as self-gravitating stellar
systems. In general, the associated two-integrals Jeans equations
(e.g., BT, Ciotti 2000) can be solved numerically
after having fixed the relative amount of ordered streaming motions
and velocity dispersion in the azimuthal direction (e.g., see Satoh
1980, Ciotti \& Pellegrini 1996). For simplicity here we restrict to
the investigation of the isotropic case, when the resulting Jeans
equations are formally identical to the equations describing
axysimmetric, self-gravitating gaseous systems in permanent rotation:
\[
\left\{
\begin{array}{lcl}
\displaystyle {1\over\rho}\, {\partial P\over\partial z} & = & 
\displaystyle -{\partial\Phi\over\partial z} \: ,\\
&&\\
\displaystyle {1\over\rho}\, {\partial P\over\partial R} & = & 
\displaystyle -{\partial\Phi\over\partial R} + \Omega^2 R ,
\end{array}
\right.
\label{eq.hydro}
\]
(in order to simplify the notation, in the following we intend
$\rho=\Re[\rhoc]$ and $\Phi=\Re[\Phic]$). The quantities $\rho$, $P$
and $\Omega$ denote the fluid density, pressure and angular velocity,
respectively; the rotational (i.e., streaming) velocity is
$\vel=\Omega R$, while $v_R =v_z =0$.

In problems where $\rho$ and $\phi$ are assigned, the standard
approach for the solution of equations~(\ref{eq.hydro}) is to
integrate for the pressure with boundary condition $P(R,\infty)=0$,
\[ 
P=\int_z^{\infty}\rho {\partial\Phi\over \partial z'} \, {\rm d} z',
\label{pressione}
\]
and then to obtain the rotational velocity field from the radial
equation
\[ 
\velq={R\over\rho}{\partial P\over\partial R} + 
           R{\partial\Phi\over\partial R}.
\label{velocita}
\]
However, $\velq$ can be also obtained without previous knowledge of
$P$, because by combining equations~(\ref{pressione})-(\ref{velocita})
and integrating by parts with the assumption that
$P=\rho\partial\Phi/\partial R=0$ for $z=\infty$ it follows that
\[ 
{\rho\velq\over R}=\int_z^{\infty}\left( 
                   {\partial\rho\over\partial R}
                   {\partial\Phi\over\partial z'}- 
		   {\partial\rho\over\partial z'} 
		   {\partial\Phi\over\partial R} 
                   \right)\, {\rm d}z'.
\label{commutatore}
\]
We remark that this ``commutator-like'' relation is not new (e.g., see
Rosseland 1926, Waxmann 1978, and, in the context of stellar dynamics,
Hunter 1977).

Before solving equations~(\ref{eq.hydro}) for the real parts of the
density-potential pairs of Sect. 2, it is useful to recall some basic
property of rotating fluid configurations. For example, in several
astrophysical application (e.g., the set-up of initial conditions for
hydrodynamical simulations), equations~(\ref{eq.hydro}) are solved for
the density under the assumption of a barotropic pressure distribution
(and neglecting the gas self-gravity), and for assigned
$\rho(R,\infty)=0$ or $\rho(R,0)=\rho_0(R)$.
%Thus, a gravitational potential $\Phi$ (for example produced by a
%stellar disk or a dark matter halo) and a specific function $P(\rho)$
%are fixed, the first of equations~(\ref{eq.hydro}) is integrated with
%boundary $\rho(R,0)$ or imposing $\rho(R,\infty)=0$, and finally the
%angular velocity $\Omega$ is obtained from the second of
%equations~(\ref{eq.hydro}).

As well known, this approach can lead only to hydrostatic ($\Omega
=0$) or {\it cylindrical rotation} ($\Omega =\Omega[R]$) kinematical
fields. In fact, according to the Poincar\'e-Wavre theorem (e.g., see
Lebovitz 1967, Tassoul 1980), cylindrical rotation is {\it equivalent}
to barotropicity, or to the fact that the acceleration field at the
r.h.s. of equations~(\ref{eq.hydro}) derives from the effective
potential
\[
\phieff\equiv\Phi - \int_{R_0}^R\Omega^2(R')R'dR' 
\] 
(where $R_0$ is an arbitrary but fixed radius), so that the gas
density and pressure are stratified on $\phieff$.

However, barotropic equilibria are just a very special class of
solutions of equations~(\ref{eq.hydro}).  For example, in the context
of galactic dynamics isotropic axisymmetric galaxy models often show
streaming velocities dependent on $z$ (for simple examples see, e.g.,
Lanzoni \& Ciotti 2003, Ciotti \& Bertin 2005). Such {\it baroclinic}
configurations (i.e., fluid systems in which $P$ cannot be expressed
as a function of $\rho$ only, and $\phieff$ does not exist at all),
have been studied in the past for problems ranging from geophysics, to
the theory of sunspots and stellar rotation (e.g., see Rosseland
1926), and to the problem of modeling the decrease of rotational
velocity of the extraplanar gas in disk galaxies for increassing $z$
(e.g., see Barnab\'e et al. 2005, 2006).  Note that a major problem
posed by the construction of baroclinic solutions is the fact that the
existence of {\it physically acceptable solutions}
(i.e. configurations for which $\velq\geq 0$ everywhere) is not
guaranteed for arbitrary choices of $\rho$ and $\Phi$. However, the
positivity of the integrand in equation~(\ref{commutatore}) for $z\geq
0$ is a sufficient condition to have $\velq \geq 0$ everywhere.  In
fact, in Barnab\'e et al. (2006) several theorems on baroclinic
configurations have been proved starting from
equation~(\ref{commutatore}): in particular {\it toroidal} gas
distributions are strongly favoured in order to have $\velq\geq 0$ in
presence of a dominating disk gravitational field.  We note that also
in the present cases the density is of toroidal shape, altough the
distribution is self-gravitating.
%%%%%%%%%%%%%%%%%%%%%%%%%%%%%%%%%%%%%%%%%%%%%%%
\begin{figure*}
\vskip -0.7truecm
\hskip -1.7truecm
\includegraphics[height=0.7\textheight,width=0.9\textwidth]{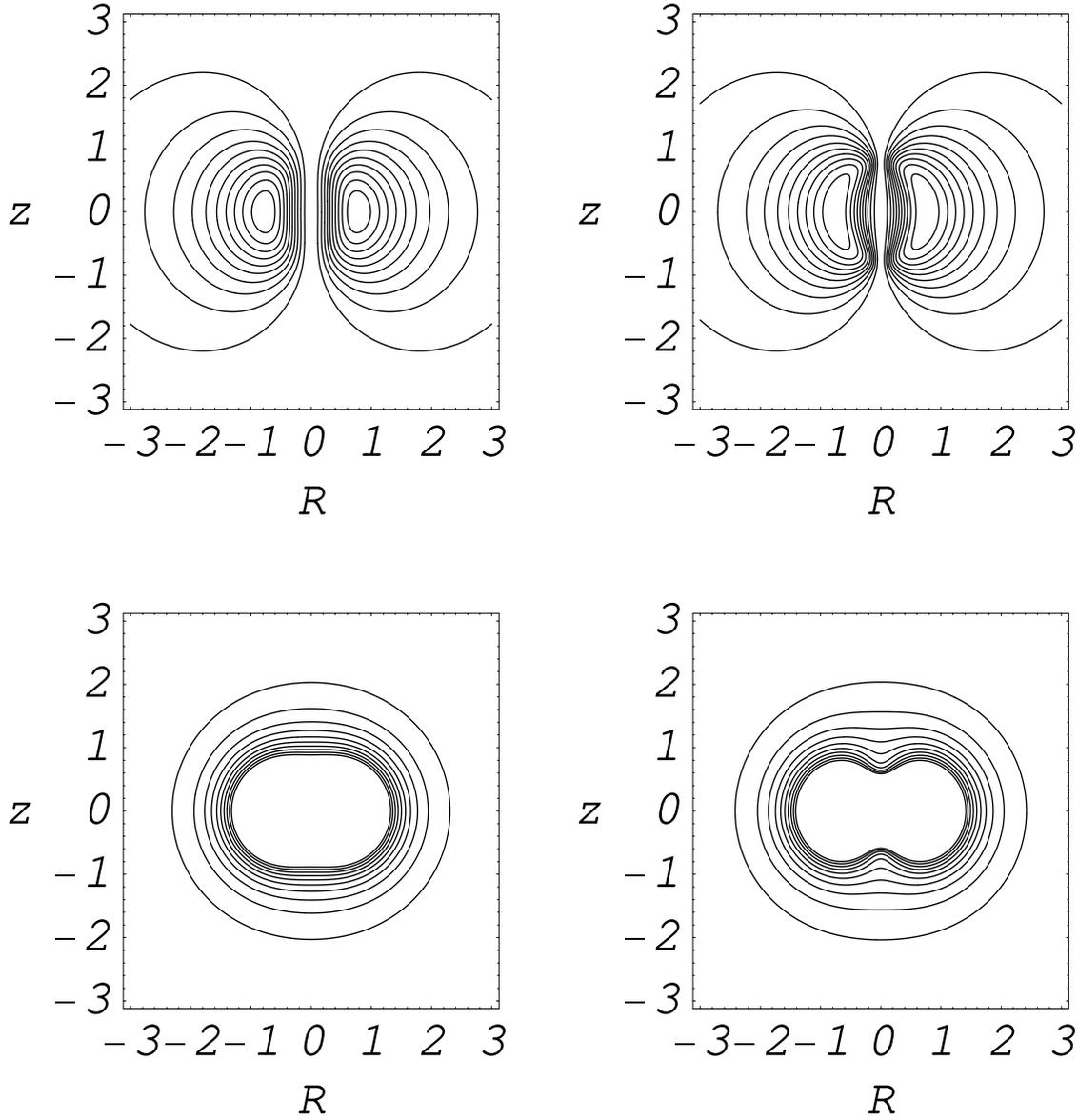}
\caption{Isorotational contours in the meridional plane of
$\Re(\rho_c)$ of Plummer sphere for $a=1/2$ (top left), and $a=23/40$
(top right).  In the bottom panels contours of constant pressure for
the same models.}
\label{plumrot}
\end{figure*}
%%%%%%%%%%%%%%%%%%%%%%%%%%%%%%%%%%%%%%%%%%%%%%%

In the case of the $\Re(\rhoc)-\Re(\Psic)$ pair of the shifted Plummer
sphere, the (normalized) commutator in equation~(\ref{commutatore}) is
a very simple function
\[
{\partial\rho\over\partial R}{\partial\Phi\over\partial z'}-
{\partial\rho\over\partial z'}{\partial\Phi\over\partial R}=
{15a^2\over\pi} {R z (1-a^2+r^2)\over d^7},
\label{complum}
\]
which is positive for $z>0$ and $0\leq a <1$, so that $\velq\geq 0$ as
far as $\Re(\rhoc)$ is positive everywhere, i.e., for $0\leq a<a_m$.
Remarkably, the integration of equation~(\ref{commutatore}) can be
easily carried out in terms of elementary functions
\begin{eqnarray}
&&\rho\velq={3 a^2 R^2\over 2\pi d^5}\Bigg\{1 -
          {d^4 (1+a^2+r^2)\over 12a^4(1+R^2)^3}\times\nonumber\\
&&
          \left[{6a^4(1+R^2)^2\over d^4}-
                {2a^2(1+R^2)\over d^2}+{1+a^2+r^2-d\over 1+a^2+r^2}
          \right]
\Bigg\},
\end{eqnarray}
while the direct integration of the pressure equation is much more
complicate (again showing the relevance of
equation~[\ref{commutatore}] in applications), and we report its
normalized expression on the equatorial plane only
\[
P(R,0)={2(1+R^2)-a^2\over 16\pi (1+R^2)^2(1-a^2+R^2)^2}.
\]
Note that $\velq\propto a^2$, and so it vanishes (as a consequence of
the imposed isotropy) when reducing to the spherically symmetric seed
density: in fact, it can be proved that the argument in parenthesis in
equation (40) is regular for $a\to 0$. The baroclinic nature of the
equilibrium is apparent, and confirmed by Fig. 2, where we show the
contours of constant $\velq$ for $a=1/2$ and $a=23/40$ (top panels,
correspondig to the models plotted in the top panels of Fig. 1).  The
normalized
streaming velocity field in the equatorial plane is
\[
\velq(R,0)={a^2 R^2\over 3}
           {6(1+R^2)^2 - 4a^2(1+R^2) +a^4\over 
           (1+R^2)^3(1-a^2+R^2)^{3/2}},
\]
and has a maximum at $R\lsim 1$, while the normalized circular
velocity is the same as that of a Plummer sphere of scale-lenght
$\sqrt{1-a^2}$ (cf. the comment after equation [26]), i.e.,
\[
v_c^2(R)=R{\partial\Phi(R,0)\over\partial R}={R^2\over (1-a^2+R^2)^{3/2}}.
\]
It is easy to verify that equations (41) and (42) satisfy equation
(36).

Unfortunately, we were unable to obtain the explicit formulae for the
analogous quantities for the shifted Isochrone model.  However, it is
easy to prove that also in this case the commutator is proportional to
$a^2$, and so it vanishes (as expected) for $a\to 0$.  A comparison of
the rotational fields of the two models in the far field can be
obtained by using the asymptotic expressions (25)-(26) and (32)-(33)
can be used.  In particular, for the real part of the shifted Plummer
sphere we found
\[
\velq={2a^2R^2\over r^5}-{5a^2 R^2(3-a^2-9a^2\mu^2)\over 3 r^7}+
      O(R^2r^{-9}),
\]
while for $\Re(\rhoc)$ of the Isochrone shifted sphere
\begin{eqnarray}
\velq&=&{a^2R^2\over r^5}
        \left({4\over 3}-{13\over 5 r}\right)+\nonumber\\
      &&{a^2R^2(3159+5780a^2+21780a^2\mu^2)\over r^7}+
          O(R^2r^{-8}).
\end{eqnarray}
Note that, even though the radial asymptotic behavior of the density
is different in the two considered cases, the obtained velocity field
has the same radial dependence in the far field, and in both cases
$\velq$ decreases for increasing $z$ and fixed $R$.  In addition,
$\velq\to 0$ for $r\to\infty$, as expected from the spherical symmetry
of the density and potential for $r\to\infty$: in fact, it is apparent
from equation~(\ref{commutatore}) that spherical (additive) components
of $\rho$ and $\Phi$ (e.g., the leading terms in the asymptotic
expansions [25]-[26] and [32]-[33]) mutually cancel in the commutator
evaluation.

\section{Conclusions}

In this paper we have shown that the complex-shift method, introduced
in electrodynamics by Newman, and throughly studied among others by
Carter, Lynden-Bell and Kaiser, can be extended to classical
gravitation to produce new and explicit density-potential pairs with
finite deviation from spherical symmetry.  In particular, we showed
that after separation of the real and imaginary parts of the
complexified density-potential pair, the imaginary part of the density
corresponds to a system of null total mass, while the real component
can be positive everywhere (depending on the original seed density 
distribution and the amount of the complex shift).

As simple application of the proposed method we illustrated the
properties of the axysimmetric systems resulting from the shift of the
Plummer and Isochrone spheres.  The analysis revealed that the
obtained densities are nowhere negative only for shift values in some
restricted range. This property is linked to the presence of a flat
"core" in the adopted (spherically symmetric) seed distributions. It
is thus expected that also other distributions, such as the King
(1972) model, and the Dehenen (1993) and Tremaine et al. (1994)
$\gamma=0$ model, will lead to physically acceptable shifted systems.
For the two new exact density-potential pairs we also found that the
associated rotational fields (obtained under the assumption of global
isotropy of the velocity dispersion tensor) correspond to baroclinic
gaseous configurations.

An interesting issue raised by the present analysis is the toroidal
shape of the obtained densities. While we were not able to find a
general explanation of this phenomenon, some hints can be obtained by
considering the behavior of the complex-shift for ${\bf a}\to 0$. In
fact, in this case it can be easily proved that while the {\it odd}
terms of the expansion correspond to the imaginary part fo the shifted
density (and contain the factor $z$), the {\it even} terms are real
and, to the second order in $a$,
\[
\Re(\rhoc)\sim\rho-{a^2\over 2r}{d\rho\over dr} + 
              {a^2 z^2\over 2 r^3}
              \left ({d\rho\over dr}-r{d^2\rho\over dr^2}\right) + O(a^4).  
\]
This expansion is functionally similar to that obtained by Ciotti \&
Bertin (2005) for the case of oblate homeoidal distributions, where
the term dependent on $z^2$ is immediately identified with a spherical
harmonic (e.g., Jackson 1999) responsible of the toroidal shape.  It
is clear that when restricting to the small-shift approximation (46),
all the computations presented in this paper can be carried out
explicitly for simple seed distributions.  Moreover, from equation
(46) it is easy to prove that negative values of density appears on
the $z$ axis - no matter how much the shift parameter is small - when
the seed density has a central cusp $\rho\propto r^{-\gamma}$ with
$\gamma >0$ (e.g., see the explicit example in Appendix).

In this paper we restricted for simplicity to spherically simmetric
seed systems, which produce axysimmetric systems. A natural extension
of the present investigation would be the study of the
density-potential pairs originated by the complex shift of seed disk
distributions, such as the Miyamoto \& Nagay (1975) and Satoh (1980)
disks, or even the generalization of homeoidal quadrature formulae
(e.g., see Kellogg 1953, Chandrasekhar 1969, BT).  Note that in such
cases, at variance with the spherical cases here discussed, the shift
{\it direction} is important: for example, the complex shift of a disk
along the $z$ axis will lead to a different system than 
an equatorial shift, that would produce a triaxial object.

A second interesting line of study could be the use of the
complex-shift to produce more elaborate density-potential pairs by
adding models with a weight function $w(\av)$, i.e., by considering
the behavior of the linear operator
\[
\varrho_c=\int\rho(\xv -i\av) w(\av) d^3\av,
\]
where the integration is extended over some suitably chosen region.
We do not pursue this line of investigation, but it is obvious that
also shifted density components with negative regions can be accepted
in this context, as far as one is able to construct a weight function
leading to a final distribution nowhere negative (as it happens, for
example, in spherical harmonics expansions).

\section*{Acknowledgments}

The paper is dedicated to the memory of my best friend and colleague
Dr. Giacomo Giampieri, who tragically passed away on September 3, 2006
while working on this project.  L.C. thanks Donald Lynden-Bell for
enligthening discussions, and Tim de Zeeuw, Scott Tremaine and an
anonymous Referee for interesting comments.  This work was supported
by the Italian MIUR grant CoFin2004 "Collective phenomena in the
dynamics of galaxies".

\appendix

\section{The shifted singular isothermal sphere}

In this case the normalized density and potential are $\rho=1/r^2$ and
$\Phi=4\pi\ln r$, so that from equations~(\ref{eqpsic})-(\ref{eqrhoc}) 
\[
\rhoc={r^2-a^2\over (r^2-a^2)^2 + 4 a^2 z^2}+ i{2az\over
      (r^2-a^2)^2 + 4 a^2 z^2},
\]
\[
\Phic=2\pi\ln (r^2-a^2 -2iaz).
\]
Note how, at variance with the Plummer and Isochrone spheres, in the
present case $\Re(\rhoc)$ describes a density distribution negative
inside the open ball $r<a$ and positive outside. The density on the
surface $r=a$ is zero except on the singular ring $R=a$ in the
equatorial plane. In the far field the density decreases as $1/r^2$,
while near the center it flattens to $-1/a^2$, while the potential
generated by $\Re(\rhoc)$ is given by
\[
\Re(\Phic)=\pi\ln [(r^2-a^2)^2 +4a^2z^2].
\]
As for the other two cases discussed in this paper, the density
$\Im(\rhoc)$ vanishes on the equatorial plane, is positive above and
negative below.

\end{document}